\documentclass[superscriptaddress,twocolumn,prl,preprintnumbers,amsmath,amssymb]{revtex4-2}
\usepackage{graphicx,afterpage}
\usepackage{xcolor}
\usepackage[colorlinks=true,plainpages=false,linkcolor=blue,urlcolor=blue,citecolor=blue,pdfpagemode=UseNone,pdfstartview=FitBH]{hyperref}

\begin{document}
\title{Singular spin fluctuations in the strange-metal phase of La$_{2-x}$Sr$_x$CuO$_4$}

\author{Benjamin Costarella} \affiliation{Univ. Grenoble Alpes, CNRS, Univ. Toulouse, INSA-T, LNCMI-EMFL, UPR3228, Grenoble, France}

\author{Lo\"ic Soriano} \affiliation{Univ. Grenoble Alpes, CNRS, Univ. Toulouse, INSA-T, LNCMI-EMFL, UPR3228, Grenoble, France}

\author{Igor Vinograd} \affiliation{Univ. Grenoble Alpes, CNRS, Univ. Toulouse, INSA-T, LNCMI-EMFL, UPR3228, Grenoble, France}

\author{Hadrien~Mayaffre} \affiliation{Univ. Grenoble Alpes, CNRS, Univ. Toulouse, INSA-T, LNCMI-EMFL, UPR3228, Grenoble, France}

\author{Shuo Li} \affiliation{Institute of Physics, Chinese Academy of Sciences, and Beijing National Laboratory for Condensed Matter Physics, Beijing 100190, China}
\author{ Jie Yang} \affiliation{Institute of Physics, Chinese Academy of Sciences, and Beijing National Laboratory for Condensed Matter Physics, Beijing 100190, China}
\author{Jun Luo} \affiliation{Institute of Physics, Chinese Academy of Sciences, and Beijing National Laboratory for Condensed Matter Physics, Beijing 100190, China}
\author{Rui Zhou} \affiliation{Institute of Physics, Chinese Academy of Sciences, and Beijing National Laboratory for Condensed Matter Physics, Beijing 100190, China} \affiliation{School of Physical Sciences, University of Chinese Academy of Sciences, Beijing 100190, China}

\author{Genda Gu}
\affiliation{Condensed Matter Physics and Materials Science Division, Brookhaven National Laboratory, Upton, New York 11973-5000, USA}

\author{John M. Tranquada}
\affiliation{Condensed Matter Physics and Materials Science Division, Brookhaven National Laboratory, Upton, New York 11973-5000, USA}

\author{Marc-Henri~Julien}
\email{marc-henri.julien@lncmi.cnrs.fr}
\affiliation{Univ. Grenoble Alpes, CNRS, Univ. Toulouse, INSA-T, LNCMI-EMFL, UPR3228, Grenoble, France}
\date{\today}

\begin{abstract}

Although spin fluctuations are central to the physics of high-$T_c$ cuprates, their relevance to strange-metal behavior in the overdoped regime remains unclear. Here, we use high magnetic fields to suppress superconductivity and an NMR protocol tailored to electronic inhomogeneity to show that the low-energy limit of the dynamical spin susceptibility $\chi''(q,\omega)$ at $x=0.25$ in La$_{2-x}$Sr$_{x}$CuO$_4$ increases continuously down to our lowest temperatures. This behavior is suggestive of quantum-critical fluctuations, a leading candidate mechanism for strange-metal transport, yet is observed well beyond the spin-stripe critical doping $x\simeq0.19$. Our data further reveal that the spin dynamics are spatially inhomogeneous, suggesting that nanoscale electronic inhomogeneity may underlie this apparent paradox. These observations provide new insight into the electronic state from which strange-metal behavior emerges.

\end{abstract}

\maketitle

The enigmatic pseudogap regime of the cuprate superconductors, which occupies a large portion of their phase diagram, has long dominated both theoretical and experimental investigations~\cite{Keimer2015}. It is, however, now becoming increasingly clear that the electronic properties of hole-doped cuprates outside their pseudogap regime---namely at hole-doping levels $p$ beyond $p^* \approx 0.2$---are themselves far more intricate than previously anticipated. In particular, this regime hosts an extended \emph{strange-metal} phase in which the electrical resistivity of the normal state is linear in temperature down to the lowest temperatures~\cite{Cooper2009,Taillefer2010,Legros2019,Phillips2022,Hussey2023}. The nature of the electronic ground state and associated excitations above $p^*$ therefore constitutes a central open problem.

While many ideas have been put forward to explain the linear resistivity~\cite{Phillips2010,Mitrano2018,Zaanen2019,Varma2020,Lee2021,Caprara2022,Thornton2023,Allocca2024,Bashan2024,Bashan2026,Fratini2025b}, spin fluctuations remain a leading candidate for strange-metal behavior in the cuprates~\cite{Chowdhury2022,Wu2022,Patel2023,Ma2023,Ciuchi2023,Teixeira2023,Patel2024,Fratini2025a,Patel2025,Chang2025}. This is not only because antiferromagnetic (AFM) correlations persist across the entire phase diagram~\cite{Fujiwara1991,Ohsugi1994,Yamada1998}, but also because they are intimately connected to the central properties of these materials, including stripe order~\cite{Tranquada2020}, pseudogap behavior~\cite{Zhou2025,Dai1999,Anderson2025,Frachet2020,Vinograd2022,Missiaen2025}, and possibly superconductivity itself~\cite{Julien1996a,Ofer2006,Dahm2009,LeTacon2011,Wang2022,Duffy2025}. 

Compelling evidence has long been lacking that spin fluctuations for $p\geq p^*$ can, on their own, account for strange-metal transport. In particular, the temperature dependence of the low-energy dynamical spin susceptibility $\chi''(q,\omega)$ inferred from earlier nuclear magnetic resonance (NMR) studies~\cite{Fujiwara1991,Vyaselev1992,Ohsugi1994,Julien1996b,Zheng2000,Williams2001,Zheng2005} appears too weak to account for linear resistivity at low temperatures~\cite{Ohsugi1994}.

This question is nevertheless timely to revisit. First, the temperature, energy, and doping dependence of low-energy spin fluctuations remains incompletely characterized. Even in the most extensively studied overdoped cuprate La$_{2-x}$Sr$_x$CuO$_4$ (LSCO), NMR~\cite{Ohsugi1994,Itoh1998,Baek2017,Frachet2020} and inelastic neutron-scattering~\cite{Yamada1998,Lee2000,Wakimoto2004,Wakimoto2005,Ikeuchi2018,Li2018,Li2022,Zhu2023,Radaelli2026} studies are relatively scarce above $p^*$, probing only limited regions of momentum, energy, temperature or doping space, and in some cases yielding results that are not yet fully reconciled. Very recent inelastic neutron-scattering measurements have provided new insight into this issue by reporting the temperature dependence of low-energy spin fluctuations at $p=0.22$ in LSCO~\cite{Radaelli2026}; we return to these results in the Discussion.

   \begin{figure}[t!]
\includegraphics[width=7cm]{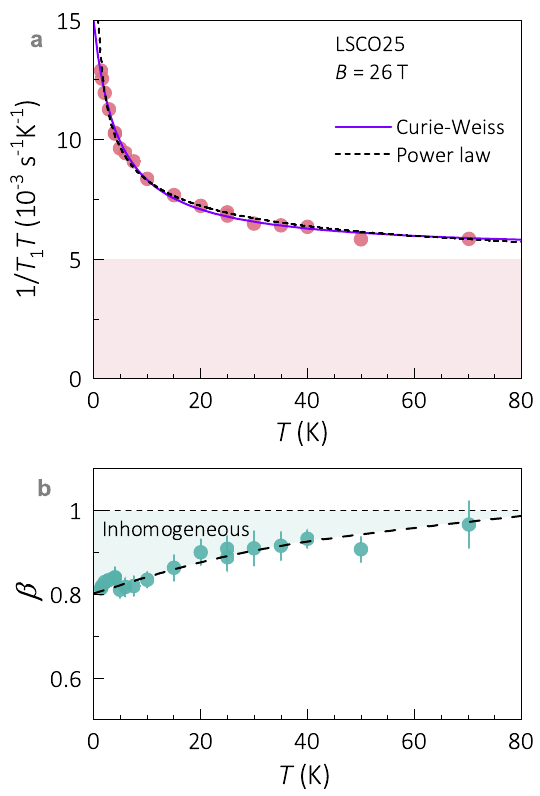}
\caption{\label{La26T} Evidence of singular behavior and spatial inhomogeneity. {\bf (a)} $^{139}$La $1/T_1T$ at 26~T. Solid and dashed lines are fits to Curie--Weiss and power-law forms, respectively, each including a constant term (shown as colored background), as described in the text. Both fits are consistent with a divergence of $\chi''(q,\omega)/\omega$ as $T \rightarrow 0$. {\bf (b)} Stretching exponent $\beta$ obtained from fits of the recovery curves (see Supplementary Information), introduced to phenomenologically account for the distribution of $T_1$ values. A homogeneous system corresponds to $\beta = 1$. Dashed lines guide the eye.
}
\end{figure} 

Second, our understanding of the cuprate normal-state has progressed considerably. It is now understood that the competition between superconductivity and stripe order restricts static spin-stripe order to the underdoped regime of LSCO, thereby hiding its connection to the pseudogap phase. Whenever superconductivity is weakened---in high magnetic fields~\cite{Frachet2020,Vinograd2022}, low-temperature-tetragonal variants~\cite{Missiaen2025,Ma2021}, and impurity-doped samples~\cite{Panagopoulos2003}---spin-stripe order persists all the way up to the pseudogap boundary $p^*$. These findings have important implications for the regime $p>p^*$: the strange metal behavior manifestly occurs on the verge of spin-stripe order---whether tuned by doping or field---suggesting that it may arise from coupling to relatively low-energy AFM fluctuations~\cite{Campbell2025}. Moreover, these developments raise the possibility that substantial low-energy AFM fluctuations in the strange-metal regime have been previously overlooked: $^{63}$Cu NMR may lose sensitivity to such fluctuations~\cite{Hunt1999,Curro2000,Julien2001,Pelc2017,Imai2021}, and they would be gapped out by superconductivity at low fields.

Here, we provide direct evidence for these hitherto hidden spin fluctuations at $p=x=0.25$ in La$_{2-x}$Sr$_x$CuO$_4$. By combining $^{139}$La (instead of $^{63}$Cu) NMR measurements with high magnetic fields $B$ to suppress superconductivity and thereby access the low-temperature normal state where strange metallic behavior occurs~\cite{Cooper2009}, we show that the low-energy dynamical spin susceptibility $\chi''(q,\omega)$ exhibits a divergence upon cooling toward zero temperature. We further discuss how the quantitative characteristics of the data constrain microscopic descriptions of the strange metal state.

   \begin{figure*}[t!]
\includegraphics[width=18cm]{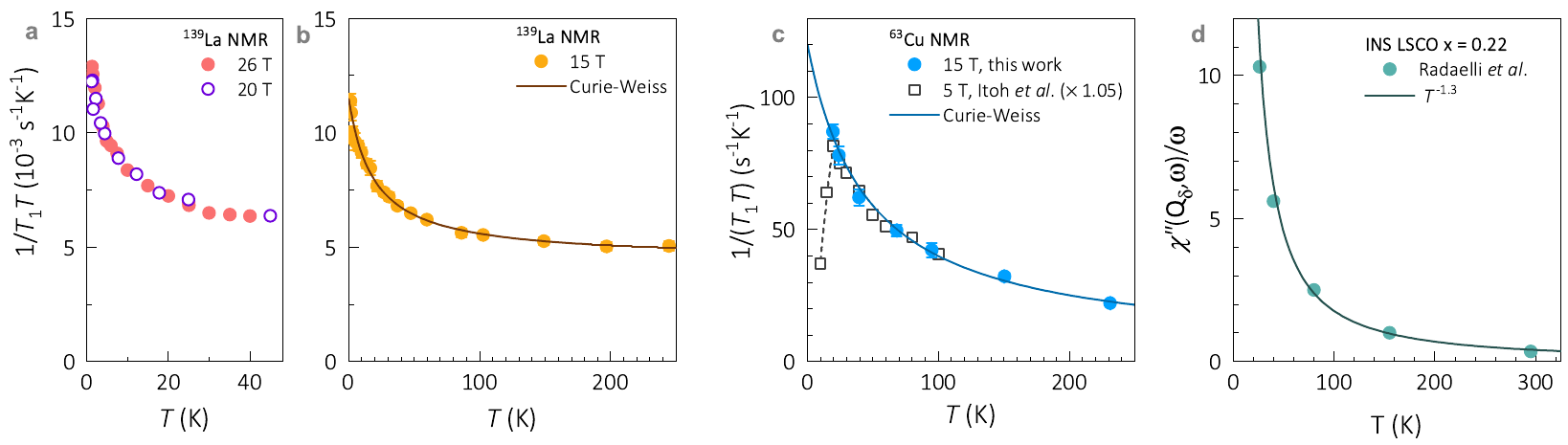}
\caption{\label{T1}
{\bf (a)} Comparison of $^{139}$La $1/T_1T$ measured at 20 and 26~T. The two data sets are nearly identical within experimental uncertainty (of the order of the symbol size). {\bf (b)} $^{139}$La $1/T_1T$ at 15~T. At this field the power-law fit becomes poor (not shown), whereas a Curie--Weiss fit (Eq.~\ref{eq:CW}) yields a Weiss temperature $\theta = -20 \pm 3$~K, significantly larger than $\theta = -6.6 \pm 2$~K at 20~T and $\theta = -4.6 \pm 0.9$~K at 26~T. {\bf (c)} $^{63}$Cu $1/T_1T$ at 15~T compared with data from Itoh {\it et al.}~\cite{Itoh1998} taken at 5~T (both with $B \parallel ab$). The low-field data show a drop at low temperature due to the opening of the superconducting gap. The solid line is a Curie--Weiss fit yielding a larger Weiss temperature $\theta = -45 \pm 23$~K and a smaller constant background ($a = 3.6$~s$^{-1}$K$^{-1}$) than obtained for $^{139}$La at the same field [panel (b)]. {\bf (d)} Temperature dependence of the slope of $\chi''(\mathbf{Q}_\delta,\omega)$ in the $\omega \rightarrow 0$ limit extracted from the neutron-scattering data of Radaelli {\it et al.}~\cite{Radaelli2026} (see Supplementary Information). Error bars correspond to two standard deviations from the fit and are not shown for clarity, as they are comparable to the symbol size.}
\end{figure*} 


{\bf Results---}Our detailed investigations show that $^{63}$Cu NMR does not provide a reliable quantitative probe of low-temperature spin dynamics in La$_{1.75}$Sr$_{0.25}$CuO$_4$ (see End Matter for sample and measurement details, as well as $^{63}$Cu data in Fig.~\ref{tau-dep}). In contrast, we demonstrate that $^{139}$La $T_1$ measurements offer a robust probe of spin fluctuations (End Matter).

To characterize the dynamical susceptibility $\chi''(q,\omega)$, we consider the quantity $1/T_1T$, which probes the low-frequency slope of $\chi''(q,\omega)$ in the $\omega \rightarrow 0$ limit ($\omega=\omega_n\simeq150$~MHz~$\sim0.6\,\mu$eV):
\begin{equation}
\frac{1}{T_1T}\propto\sum_{q,\alpha\perp B} F_{\alpha\alpha}(q)
\frac{\chi^{\prime\prime}_{\alpha\alpha}(q,\omega_n)}{\omega_n} ,
\label{eq:T1}
\end{equation}
where the sum over wave-vectors $q$ is weighted by a $q$-dependent hyperfine form factor $F(q)$.

The spin-lattice relaxation rate $1/T_1T$ of $^{139}$La was measured at $B=26$~T applied perpendicular to the CuO$_2$ planes, a field comparable to the superconducting upper critical field $B_{c2}$ at this doping level (see Refs.~\cite{Frachet2020,Girod2021}). At 26~T and $T\simeq2$~K, the response of the NMR tank circuit suggests that superconductivity remains present, albeit in a very weak form.

At 26~T, $1/T_1T$ increases monotonically upon cooling down to 1.34~K, the base temperature of our experiment, with no indication of saturation (Fig.~\ref{La26T}a). The data can be described either by a Curie--Weiss form,
\begin{equation}
\frac{1}{T_1T} = a + \frac{b}{T - \theta},
\label{eq:CW}
\end{equation}
with a very small Weiss temperature $\theta = -4.6$~K, or by a power law,
\begin{equation}
\frac{1}{T_1T} = c + d\,T^{-(1+\alpha)},
\label{eq:power}
\end{equation}
with $\alpha = -0.67$ (using the same definition of $\alpha$ as in ref.~\cite{Radaelli2026}). Both parameterizations are consistent with a divergent, or nearly divergent, $\chi''(q,\omega\rightarrow0)$ as $T \rightarrow 0$. 

$T_1(T)$ was also measured at two other magnetic fields. The data at 20~T are indistinguishable within experimental uncertainty from those at 26~T (Fig.~\ref{T1}a). Therefore, although superconductivity remains present at the lowest temperatures at 26~T, it is too weak to measurably affect $T_1$. We therefore conclude that our high-field measurements access the intrinsic normal state and that, under these conditions, $\chi''(q,\omega \rightarrow 0)$ displays a singular temperature dependence in the $T \rightarrow 0$ limit. 

At 15~T, on the other hand, the low-temperature increase is weaker (Fig.~\ref{T1}b), yielding a significantly larger Weiss temperature $\theta\simeq-20$~K. This indicates that, even though no gap is visible in $1/T_1T$, residual superconducting correlations suppress the intrinsic divergence. The full field evolution toward the zero-field limit could not be tracked due to the severe loss of NMR signal intensity in the superconducting state of single crystals. This difficulty is compounded by the fact that, even in zero field, the gap-like behavior is attenuated by a large residual $1/T_1T$~\cite{Ohsugi1994}, indicating that a significant fraction of carriers remains uncondensed.

\begin{figure*}[b!]
\includegraphics[width=18cm]{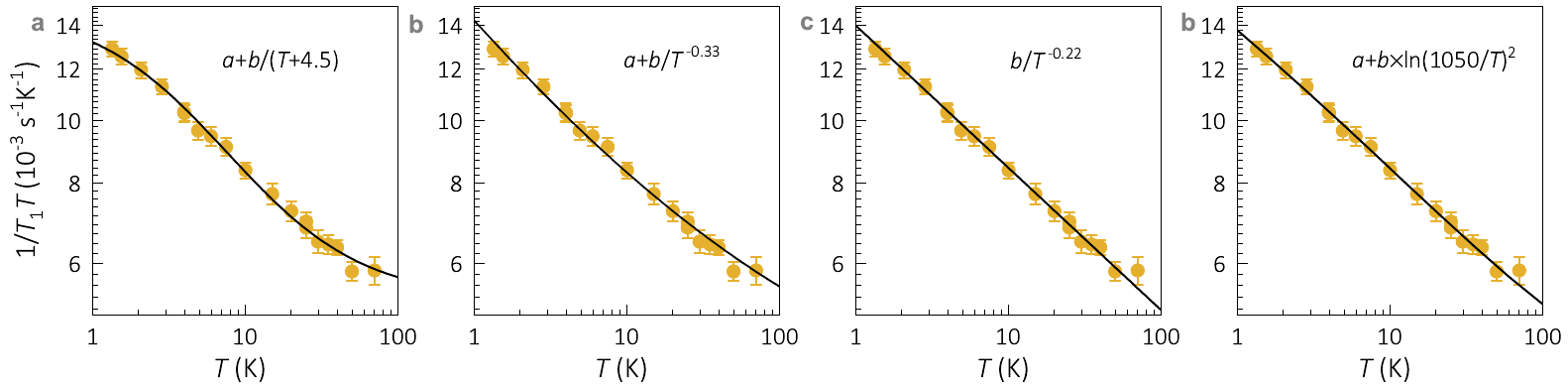}
\caption{\label{4fits} Various fits, all of comparable quality, to the temperature dependence of $^{139}$La $1/T_1T$ at 26~T.}
\end{figure*}

We note that our $1/T_1T$ data can also be fit by the form $a + b\,[\ln(T_0/T)]^2$ (Fig.~\ref{4fits}d), as expected in the vicinity of a two-dimensional Van Hove singularity---known to cross the Fermi level ($E_F$) at $p=0.21$ in LSCO~\cite{Zhong2022}. However, the temperature-induced variation of the density of states (DOS) at $E_F$ inferred from our data corresponds to that calculated at doping $p=0.22$, based on the experimental three-dimensional band structure of LSCO~\cite{Zhong2022}, whereas the calculation for $p=0.24$ shows essentially no temperature dependence. This indicates that a DOS effect cannot account for our observations at $p=0.25$, even allowing for the presence of regions with local doping $p=0.22$ within the sample~\cite{Li2022}. This conclusion is further supported by the observation that the van Hove singularity does not manifest in the static spin susceptibility $\chi'(q=0,\omega=0)$~\cite{Nakano1994}.

 \begin{figure*}[t!]
\includegraphics[width=18cm]{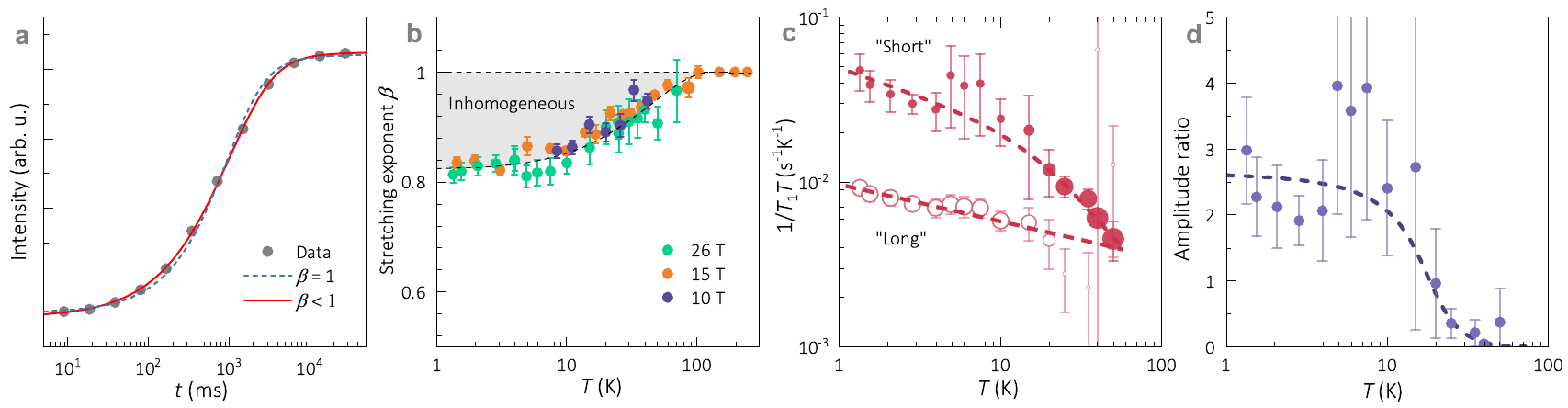}
\caption{\label{inhomogeneity} {\bf a}, $^{139}$La nuclear magnetization recovery after an inversion pulse ($T=4$~K, $B=26$~T). The better fit with the stretching exponent $\beta<1$ is evidence of a distribution of $T_1$ values. {\bf b}, Temperature dependence of the stretching exponent $\beta$ at different magnetic field values (data at 26 T are identical to those in Fig.~\ref{La26T}b). Inhomogeneity becomes noticeable below approximately 100~K and is remarkably field independent. {\bf c}, Temperature dependence of the two $T_1$ values obtained from a two-component fit of the recoveries (the fit is practically indistinguishable from the stretched fit in (a) and is thus not shown). The varying symbol size schematically reflects the varying weight of the two components, shown in (d). {\bf d}, Weight ratio of the two components. The short (more magnetic) component represent a relatively minor ($\sim$25-30\%) fraction of the sites. }
\end{figure*} 

{\bf Significance---}The significance of our observation of singular behavior in $\chi''(q,\omega \rightarrow 0)(T)$ is fourfold. First, this strong enhancement was not reported in prior NMR studies because sufficiently high magnetic fields were not used and the advantages of $^{139}$La NMR over $^{63}$Cu NMR in this regime had not yet been recognized. Second, the divergence is established here down to low temperatures in the field-induced normal state, such that spin fluctuations are probed under the same conditions in which the linear term in the resistivity is isolated, without relying on extrapolations of normal-state data below the zero-field $T_c$. Third, the singular behavior of $\chi''(q,\omega \rightarrow 0)$ is suggestive of quantum-critical dynamics, for which scaling arguments predict a power-law behavior 
$1/T_1\propto T^{-\alpha}$, with an exponent $\alpha$ determined by the critical exponents $d$, $z$ and $\eta$~\cite{Millis1993,Chubukov1993}. Fourth, such dynamics are widely considered a leading candidate mechanism for linear-in-$T$ resistivity~\cite{Phillips2022}.

These observations do not demonstrate that strange metallicity is related to quantum-critical-like spin fluctuations. However, these fluctuations are observed under the same conditions of doping, temperature and field as those in which linear resistivity emerges, which makes them a plausible microscopic origin of strange-metal transport. We note that similar behavior in $1/T_1T$ has been reported at the AFM quantum critical point (QCP) in BaFe$_2$(As$_{1-x}$P$_x$)$_2$ in conjunction with linear resistivity~\cite{Nakai2010}, as well as in ${\rm Na}_x({\rm H}_3{\rm O})_z{\rm CoO}_2\cdotp y{\rm H}_2{\rm O}$, upon suppressing superconductivity with a magnetic field~\cite{Ihara2007}.

{\bf Relation to neutron scattering---}A complementary perspective is provided by recent inelastic neutron-scattering measurements of the low-energy spin dynamics in La$_{1.78}$Sr$_{0.22}$CuO$_4$~\cite{Radaelli2026}. Extending earlier work~\cite{Zhu2023}, these authors report quantum-critical scaling of the response at the incommensurate spin-stripe wave vector $\mathbf{Q}_{\delta}$ (close to $\mathbf{Q}_{\rm AF}=(\pi,\pi)$)~\cite{Radaelli2026}. Extracting from their data the $\omega \rightarrow 0$ slope of $\chi''(\mathbf{Q}_{\delta},\omega)$ (Fig.~\ref{T1}d)---a quantity directly related to $1/T_1T$ (Eq.~\ref{eq:T1})---reveals a pronounced increase upon cooling that is consistent with the divergence inferred from our $^{139}$La measurements.

Although the neutron data extend only down to 26~K, where superconductivity sets in at zero field, the reported divergence ($\chi''(\mathbf{Q}_{\delta},\omega)/\omega \propto T^{-1.3}$ in Fig.~\ref{T1}d, consistent with $\alpha=0.32$ in~\cite{Radaelli2026}) is stronger than that inferred from NMR. This difference likely reflects both the mismatch in doping levels ($p=0.22$ vs. 0.25) and the fact that neutron scattering probes $\chi''$ near $\mathbf{Q}_{\delta}$, whereas NMR measures a weighted sum over all wave vectors (Eq.~\ref{eq:T1}).

Another notable difference between NMR and neutron scattering is the presence, in $^{139}$La $1/T_1T$, of a sizable $T$-independent contribution in addition to the singular term (Fig.~\ref{La26T}). The constant term is nearly absent in $^{63}$Cu data (Fig.~\ref{T1}c). Unlike $^{63}$Cu, the $^{139}$La form factor $F(q)$ in Eq.~\ref{eq:T1} disfavors wave vectors near $\mathbf{Q}_{\rm AF}$~\cite{Chakravarty1991}. This suggests that the constant component originates primarily from excitations away from $\mathbf{Q}_{\rm AF}$. Nonetheless, its precise interpretation remains to be clarified. It might be tempting to interpret the constant $\chi''(T)$ as a conventional Fermi-liquid-like component, as in the two-electronic-fluid picture discussed in recent works~\cite{Ayres2021,Juskus2024,Li2022,Tranquada2024}. However, a roughly constant background has also been observed for $0.07\leq p \leq 0.20$~\cite{Baek2017}, and is virtually identical to that found here at $p=0.25$, making a simple interpretation in terms of DOS at $E_F$ unlikely. 

{\bf Which criticality?---}Spin-stripe order terminates at $p^* \simeq 0.19$ in the absence of superconductivity~\cite{Frachet2020,Vinograd2022}. Therefore, the observation of a singular $\chi''(q,\omega \rightarrow 0)$ at a doping level not associated with any known QCP is difficult to reconcile with standard quantum critical scenarios. This raises the possibility of an extended doping range of critical fluctuations~\cite{Zhu2023,Radaelli2026}, and the question is whether this behavior is intrinsic or disorder-driven. For example, ref.~\cite{Prelovsek2004} predicts a form of $\chi''$ compatible with our observations---including the constant offset---even in the absence of a clearly identified QCP. On the other hand, it is difficult to overlook the fact that, in LSCO, nanoscale variations in the hole concentration are such that puddles with local doping $p_{\rm loc}\lesssim p^*$ are expected to exist far above $p^*$~\cite{Singer2002,Singer2005,Li2022}. Persistent, quantum-critical-like dynamics could then arise from the inability of spin stripes to freeze in these puddles once they become disconnected from each other above $p^*$~\cite{Tranquada2024}.

{\bf Spatial inhomogeneity---}Static and dynamic aspects of spatial inhomogeneity can be probed by lineshape and $T_1$ measurements, respectively.

The lineshape data are difficult to interpret. The $^{139}$La lines (Fig.~\ref{linewidth}a,b) are structureless, showing no evidence for the two distinct environments expected in a simple two-phase scenario. This absence of resolved features is not surprising: both dynamical averaging (due to carrier motion) and the fact that each $^{139}$La nucleus is coupled to two CuO$_2$ planes tend to average out local differences, especially if they occur on short length scales. The $^{139}$La central line nevertheless broadens upon cooling (Fig.~\ref{linewidth}c), signaling increasing inhomogeneity, although this effect is not specific and may arise from hole-doping variations as well as from spin- and charge-density oscillations around defects, as commonly observed in correlated materials.

The $T_1$ data, by contrast, directly reveal spatially inhomogeneous spin dynamics: the recovery of the nuclear magnetization deviates from the exponential form expected for a single $T_1$ (Fig.~\ref{inhomogeneity}a). The distribution of $T_1$ values is captured phenomenologically by a stretching exponent $\beta$ (Fig.~\ref{inhomogeneity}a and End Matter), and the values shown in Fig.~\ref{La26T} actually correspond to the median of this distribution. $\beta$ progressively deviates from unity already above $T_c$ (below $\sim 100$~K), reaching a field-independent value at low temperature (Figs.~\ref{La26T}b and~\ref{inhomogeneity}b). The dynamical inhomogeneity is thus an intrinsic property of the low-temperature normal state.

To further probe the two-component scenario, we performed an alternative analysis using two distinct (unstretched) $T_1$ components. These fits are of comparable quality to the stretched fits and yield two relaxation rates differing by a factor of about five at low temperature (Fig.~\ref{inhomogeneity}c). Both components increase upon cooling, indicating a continuous distribution of magnetic fluctuations rather than a sharp separation into regions with and without critical dynamics. The data also rule out spin freezing in part of the sample. The faster component accounts for roughly 25--30\% of the sample volume (Fig.~\ref{inhomogeneity}d), consistent with the expected $\sim 20$\% of sites located in striped regions with local doping $p\leq p^*$~\cite{Li2022,Tranquada2024}. That these regions occupy only a minority of lattice sites is also compatible with a quantum Griffiths phase~\cite{Patel2023,Patel2024,Patel2025,Radaelli2026}. There is in fact no clear distinction between these two scenarios, as doping inhomogeneity likely provides a microscopic realization of the strong randomness invoked in quantum Griffiths descriptions. The importance of disorder and spatial inhomogeneity in this region of the phase diagram is further highlighted by other theoretical works~\cite{Li2021,Sulangi2025}. Our analysis, however, remains subject to significant uncertainty, and further progress will require a direct determination of the $T_1$ distribution, which remains an important challenge~\cite{Arsenault2020}.

{\bf Outlook---}In conclusion, our observation of singular low-energy spin fluctuations and spatial inhomogeneity in the strange-metal phase of La$_{2-x}$Sr$_x$CuO$_4$ provides quantitative benchmarks for microscopic descriptions of this phase. These results highlight the need for a more detailed characterization of spatial inhomogeneity and open clear perspectives for systematic studies across doping levels and cuprate families.

{\bf Acknowledgments---}Work at LNCMI was supported by the Laboratoire d'Excellence LANEF (ANR-10-LABX-51-01) and by the French Agence Nationale de la Recherche (ANR) under reference ANR-25-CE30-2817 (Strangemetal). Part of this work was performed at the Laboratoire National des Champs Magn\'etiques Intenses, a member of the European Magnetic Field Laboratory (EMFL). A portion of this work was carried out at the Synergetic Extreme Condition User Facility (SECUF, https://cstr.cn/31123.02.SECUF). Work at Brookhaven is supported by the Office of Basic Energy Sciences, Materials Sciences and Engineering Division, U.S. Department of Energy (DOE) under Contract No. DE-SC0012704.

\clearpage

\section*{End Matter}

\begin{figure*}[t!]
\includegraphics[width=15cm]{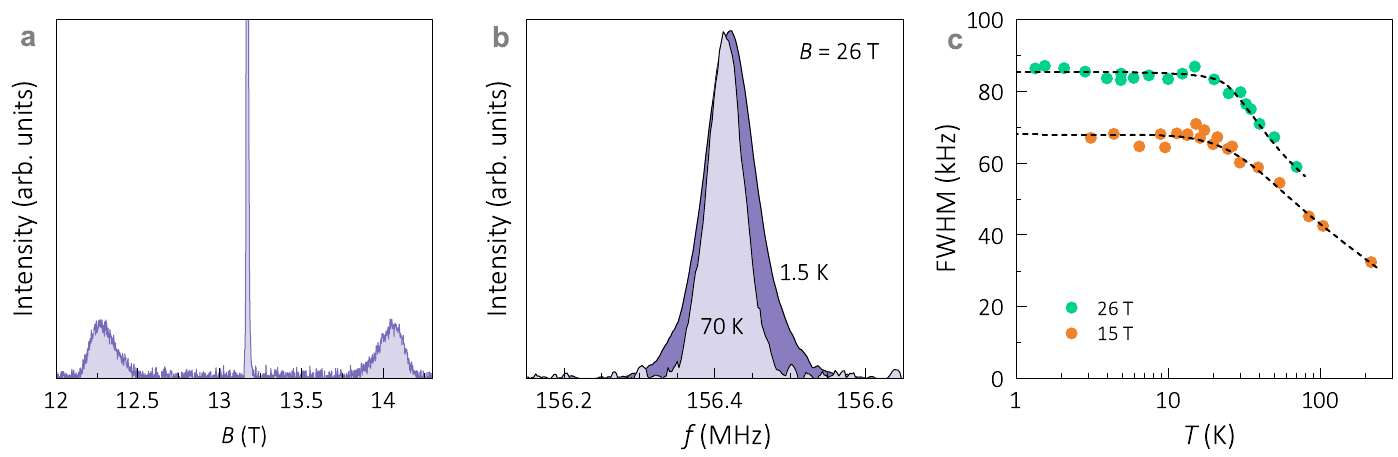}
\caption{\label{linewidth} $^{139}$La NMR characterization of static inhomogeneity. {\bf a}, Central line and first quadrupole satellites as a function of field , and $T=30$~K. The full-width-at-half-maximum (FWHM) of the first satellites is $955\pm20$~kHz. {\bf b}, Central line at 1.5 and 70~K, and $B=26$~T. {\bf c}, Temperature dependence of the FWHM of the central line, at 15 and 26 T. Because of mixed quadrupolar and magnetic hyperfine contributions, the FWHM does not simply scale with either $B$ or $1/B$. Dashed lines guide the eye. The thermal evolution of this static inhomogeneity do not show any correlation with that of the $T_1$ inhomogeneity (see text). }
\end{figure*}

{\bf Methods---}The La$_{1.75}$Sr$_{0.25}$CuO$_4$ single crystal is from the same batch as that studied in Refs.~\cite{Li2022,Tranquada2024} and has an identical bulk superconducting transition temperature $T_c=18$~K. We refer the reader to these references for details of the synthesis and sample characterization. The widths of both central and satellite lines (Figs.~\ref{linewidth}a--c) are consistent with literature data and with our own unpublished measurements in other high-quality single crystals.

All experiments were performed in superconducting magnets, including the 26~T magnet at the Synergetic Extreme Conditions User Facility (SECUF) in Beijing. The field was applied perpendicular to the $ab$ planes, unless specified otherwise. 

$T_1$ was measured using an inversion-recovery sequence. The $T_1$ values were determined by fitting the recovery curves $M(t)$ to a stretched version of the theoretical expression for magnetic relaxation between the $m_I=\pm 1/2$ levels. For $^{139}$La (nuclear spin $I=7/2$), this reads:
\begin{multline}
M(t) =  M_0 \Big[ 1-a\Big(0.714 \, e^{-\left({28t \over T_1}\right)^\beta} - \, 0.206\, e^{-\left({15t \over
T_1}\right)^\beta}\\
 - \, 0.068 \, e^{-\left({6t \over T_1}\right)^\beta}  - \,0.012\, e^{-\left({t \over T_1}\right)^\beta} \Big)\Big] ,
\label{recovery}
\end{multline}
where $M_0$ (the equilibrium nuclear magnetization) and $a$ ($a=2$ for a perfect inversion) were fit parameters, together with $T_1$ and the stretching exponent $\beta$.

{\bf Limitations of $^{63}$Cu NMR---}In a typical experiment, a spin-echo signal is recorded at a time $\tau$ following a sequence of radio-frequency (r.f.) pulses. During this delay, the signal decays with a characteristic timescale known as the spin-spin relaxation time $T_2$. If $T_2$ is spatially homogeneous, the choice of $\tau$ affects only the overall signal intensity. By contrast, when $T_2$ becomes too short for the signal to be detected in parts of the sample, the remaining signal at finite $\tau$ is no longer statistically representative of the entire sample and the measured response can acquire an explicit $\tau$ dependence. 

Such gradual ‘wipeout’ of the $^{63}$Cu NMR signal has been known to arise from the glassy freezing of spin stripes, for $p \leq 0.14$ in LSCO~\cite{Hunt1999,Curro2000,Julien2001}. However, the recent discovery that spin freezing persists up to $p\simeq0.19$ in high fields~\cite{Frachet2020,Vinograd2022} implies that $^{63}$Cu signal wipeout should also occur in these conditions. It is even conceivable that wipeout persists beyond $p^*$ if low-energy spin-stripe fluctuations remain significant despite the loss of order. 

\begin{figure*}[t]
\includegraphics[width=14cm]{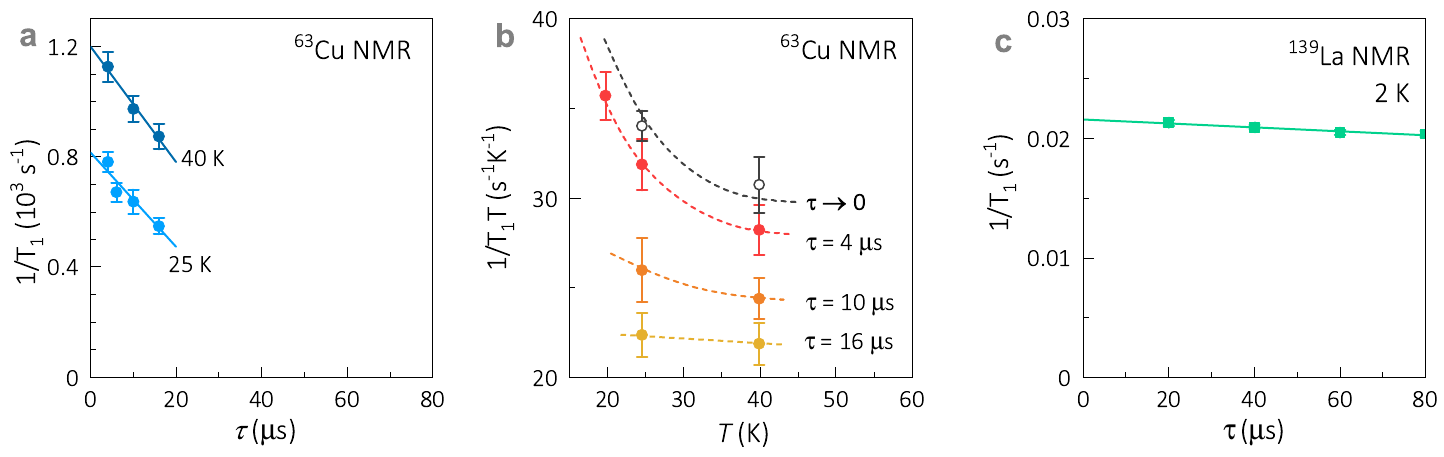}
\caption{\label{tau-dep}Effect of fast, inhomogeneous $T_2$ on $^{63}$Cu and $^{139}$La $T_1$ measurements. {\bf (a)} Dependence of $^{63}$Cu $1/T_1$ on the delay time $\tau$ between the two radio-frequency pulses of the NMR spin-echo sequence, measured at 40 and 25~K. {\bf (b)} Temperature dependence of $^{63}$Cu $1/T_1T$ for different values of $\tau$. Open symbols correspond to the $\tau \rightarrow 0$ limit of the data in panel (a). For $\tau=16$~$\mu$s, $1/T_1T$ stays flat between 40 and 25~K---similar to the data of ref.~\cite{Ohsugi1994}---but increases upon cooling for $\tau=4$~$\mu$s as in the data of ref.~\cite{Itoh1998}. The average $T_2$ becomes so short that the NMR signal eventually disappears below 19~K, even for $\tau$ as short as 4~$\mu$s---the minimum accessible with our spectrometer. {\bf (c)} Dependence of $^{139}$La $1/T_1$ on $\tau$. In contrast to $^{63}$Cu, the $\tau$ dependence is negligible.}
\end{figure*}

Our $^{63}$Cu NMR results fully validate this hypothesis: while we cannot directly measure the signal intensity because it is limited by the skin depth that varies with $T$, we find that not only the measured value of $T_1$ depends on $\tau$ (Fig.~\ref{tau-dep}a) but, even more problematically,  so does its $T$ dependence (Fig.~\ref{tau-dep}b). We emphasize that extensive tests ruled out transient heating by r.f. pulses as the origin of the $\tau$ dependence of the results. These findings show that the contrasting $T$ dependences of $1/T_1$ between refs.~\cite{Ohsugi1994} (nuclear quadrupole resonance for $x=0.24$) and \cite{Itoh1998} (NMR for $x=0.23$ with $B$ applied parallel to the $ab$ planes) arise neither from a doping difference nor from an effect of the magnetic field but from the use of different $\tau$ values in these two experiments. 

A workaround to the $\tau$ dependence could be to extrapolate the $T_1$ values at $\tau=0$ (Fig.~\ref{tau-dep}a,b). However, this would not solve another problem: $T_2$ shortens so much that the $^{63}$Cu signal is completely lost below $\approx 19$~K. $^{63}$Cu NMR thus fails to faithfully capture the spin dynamics in LSCO, not only below $p^*$ but also above.

{\bf Robustness of $^{139}$La NMR---}To circumvent this problem, we turn to another nucleus with longer $T_2$. As extensive studies for $p<p^*$ have demonstrated, $^{139}$La, with its weak but finite hyperfine coupling, is well suited for this purpose. We find that the $\tau$ dependence of $T_1$ for $^{139}$La (Fig.~\ref{tau-dep}c) is extremely small compared to that for $^{63}$Cu: the difference between $T_1$ measured at short $\tau$ (2--10~$\mu$s) and $T_1$ extrapolated to $\tau=0$ is negligible. We also find that $T_1$ for $^{139}$La is approximately four thousand times longer than for $^{63}$Cu, consistent with a relaxation mechanism dominated by spin fluctuations rather than by charge or lattice fluctuations. 


\end{document}